\PassOptionsToPackage{caption}{style=base}
\documentclass[notitlepage,nofootinbib,11pt]{revtex4-2}
\usepackage{silence}
\usepackage{amsmath,amssymb,amsfonts,mathtools}
\usepackage[dvipsnames]{xcolor}
\usepackage[toc,page]{appendix}
\usepackage{csl}
\usepackage{hyperref,dsfont}
\hypersetup{
    unicode=false,          
    pdftoolbar=true,        
    pdfmenubar=true,        
    pdffitwindow=false,     
    pdfstartview={FitH},    
    pdftitle={My title},    
    pdfauthor={Author},     
    pdfsubject={Subject},   
    pdfcreator={Creator},   
    pdfproducer={Producer}, 
    pdfkeywords={keyword1, key2, key3}, 
    pdfnewwindow=true,      
    colorlinks=true,       
    linkcolor=blue,          
    citecolor=blue,        
    filecolor=magenta,      
    urlcolor=cyan,           
}
\DeclareMathAlphabet{\mathpzc}{OT1}{pzc}{m}{it}

\def\!#1{\mathcal{#1}}
\def\*#1{\boldsymbol{\mathbf{#1}}}
\def\|#1{\textnormal{#1}}
\def\##1{\mathpzc{#1}}
\renewcommand{\Re}{\mathds{R}}

\def\norm#1{\big\lVert#1\big\rVert}

\DeclareMathOperator{\tr}{tr}

\newcommand{\Ct}[2]{\widetilde{\*\Sigma}_{#1#2}}
\newcommand{\Bt}{\widetilde{\*B}}

\newcommand{\xf}{X^\|f}
\newcommand{\cxf}{\!X^\|f}
\newcommand{\bxf}{\*X^\|f}

\newcommand{\Id}{\mathbf{I}}
\newcommand{\R}{\mathbf{R}}
\renewcommand{\S}{\mathbf{S}}
\newcommand{\transp}{\mathsf{T}}
\newcommand{\Tp}{\mathsf{T}}

\usepackage{caption}
\usepackage{subcaption}

\numberwithin{equation}{section}

\graphicspath{ {Figures/} }

\makeatletter
\renewcommand{\p@subsection}{}
\renewcommand{\p@subsubsection}{}
\makeatother

\makeatletter\def\fps@figure{htbp}\makeatother

\allowdisplaybreaks[1]

\begin{document}

\csltitle{A Stochastic Covariance Shrinkage Approach in Ensemble Transform Kalman Filtering}
\cslauthor{Andrey A Popov, Adrian Sandu, Elias D. Nino-Ruiz, Geir Evensen}
\cslyear{20}
\cslreportnumber{3}
\cslemail{apopov@vt.edu, sandu@cs.vt.edu, enino@uninorte.edu.co, geev@norceresearch.no}
\csltitlepage

\title{A Stochastic Covariance Shrinkage Approach in Ensemble Transform Kalman Filtering}
\author{Andrey A Popov}
\email[]{apopov@vt.edu}
\affiliation{Computational Science Laboratory, Department of Computer Science, Virginia Tech}

\author{Adrian Sandu}
\email[]{sandu@cs.vt.edu}
\affiliation{Computational Science Laboratory, Department of Computer Science, Virginia Tech}

\author{Elias D. Nino-Ruiz}
\email[]{enino@uninorte.edu.co}
\affiliation{Applied Math and Computer Science Lab, Universidad del Norte, Colombia}

\author{Geir Evensen}
\email[]{geev@norceresearch.no}
\affiliation{Norwegian Research Center (NORCE) and Nansen Environmental and Remote Sensing Center (NERSC), Norway}
\date{\today}

\begin{abstract}
The Ensemble Kalman Filters (EnKF) employ a Monte-Carlo approach to represent covariance information, and are affected by sampling errors in operational settings where the number of model realizations is much smaller than the model state dimension. To alleviate the effects of these errors EnKF relies on model-specific heuristics such as covariance localization, which takes advantage of the spatial locality of correlations among the model variables. This work proposes an approach to alleviate sampling errors that utilizes a locally averaged-in-time dynamics of the model, described in terms of a climatological covariance of the dynamical system. We use this covariance as the target matrix in covariance shrinkage methods, and develop a stochastic covariance shrinkage approach where synthetic ensemble members are drawn to enrich both the ensemble subspace and the ensemble transformation. We additionally provide for a way in which this methodology can be localized similar to the state-of-the-art LETKF method, and that for a certain model setup, our methodology significantly outperforms it.
\end{abstract}

\maketitle

\section{Introduction}
The ensemble Kalman filter \cite{evensen1994sequential,burgers1998analysis,evensen2009data}, one of the most widely applied data assimilation  algorithms \cite{asch2016data,law2015data,reich2015probabilistic}, uses a Monte Carlo approach to provide a non-linear approximation to the Kalman filter~\cite{kalman1960new}. In the typical case of an undersampled ensemble the algorithm requires correction procedures such as inflation~\cite{anderson2001ensemble}, localization~\cite{hunt2007efficient, petrie2008localization, anderson2012localization,Sandu_2015_SCALA,Sandu_2017_Covariance-Cholesky,zhang2010ensemble}, and ensemble subspace enrichment~\cite{Sandu_2015_covarianceShrinkage, Sandu_2019_Covariance-parallel,Sandu_2014_EnKF_SMF}. 

Hybrid data assimilation \cite{hamill2000hybrid} is typically an umbrella term for assimilation techniques that combine both offline-estimated climatological covariances with their online-estimated statistical counterparts. These methods are often thought of as heuristic corrections, but in fact stem from  statistically rigorous covariance shrinkage techniques.

This work is based on enriching the ensemble subspace through the use of climatological covariances. Previous work~\cite{Sandu_2015_covarianceShrinkage, Sandu_2019_Covariance-parallel} proposed augmenting the covariance estimates derived from the ensemble by a full rank shrinkage covariance matrix approximation. In this work we consider augmenting the physical ensemble with synthetic members drawn from a normal distribution with a possibly low rank covariance matrix derived from \textit{a priori} information such a climatological information or method of snapshots. We show that this is equivalent to a stochastic implementation of the shrinkage covariance matrix estimate proposed in ~\cite{Sandu_2015_covarianceShrinkage, Sandu_2019_Covariance-parallel}, and therefore augmenting the physical ensemble with synthetic members enriches the rank of the covariance matrix, and nudges the resulting covariance estimate toward the true covariance.

\section{Background}
Our aim is to understand the behavior of an evolving natural phenomenon. The evolution of the natural phenomenon is approximated by an imperfect dynamical model:
\begin{equation}
    X_i = \!M_{i-1,  i}(X_{i-1}) + \*\xi_i,
\end{equation}
where $X_{i-1}$ is a random variable (RV) whose distribution represents our uncertainty in the state of the system at time $i-1$, $\!M_{i-1,  i}$ is the (imperfect) dynamical model, $\*\xi_i$ is a RV whose distribution represents our uncertainty in the additive modeling error, and $X_i$ is the RV whose distribution represents our uncertainty in the (forecasted) state at time $i$.

One collects noisy observations of the truth:
\begin{equation}
    \*y^\|o_i = \!H_i(\*x_i^\|t) + \*\eta_i,
\end{equation}
where $\*x^\|t$ represents the true state of nature represented in model space, $\!H_i$ is the (potentially non-linear) observation operator, $\*\eta_i$ is a RV whose distribution represents our uncertainty in the observations, and $\*y^\|o_i$ are the observation values, assumed to be realizations of an observation RV $Y_i$. Take $n$ to be the dimension of the state-space, and $m$ to be the dimension of the observation space.

The goal of data assimilation is to find the \textit{a posteriori} estimate of the state given the observations, which is typically achieved through Bayes' theorem. At time $i$ we have:
\begin{equation}
    \pi(X_i|Y_i) \propto \pi(Y_i|X_i)\,\pi(X_i).
\end{equation}
In typical Kalman filtering the assumption of Gaussianity is made, whereby the states at all times, as well as the additive model and observation errors, are assumed to be Gaussian and independently distributed. Specifically one assumes $\*\xi_i \sim  \!N(\*0, \*Q_i)$ and $\*\eta_i \sim \!N(\*0,\R_i)$. 

In what follows we use the following notation. The \textit{a priori} estimates at all times are represented with the superscript $\square^\|f$, for forecast (as from the imperfect model), and the \textit{a posteriori} estimates are represented with the superscript $\square^\|a$, for analysis (through a DA algorithm).

\subsection{Ensemble Transform Kalman Filter}
Forecasting with an ensemble of coarse models has proven to be a more robust methodology than forecasting with a single fine model~\cite{kalnay2003atmospheric}. Ensemble Kalman filtering aims to utilize the ensemble of forecasted states to construct empirical moments and use them to implement the Kalman filter formula. The Ensemble Transform Kalman Filter (ETKF) \cite{bishop2001adaptive} computes an optimal transformation of the prior ensemble member states to the posterior member states; for Gaussian distributions the optimal transform is described by a symmetric transform matrix.

We now describe the standard ETKF. Let $\*X^\|a_{i-1} = [\*x^{(1),\|a}_{i-1},\dots \*x^{(N),\|a}_{i-1}]$ represent the $N$--members analysis ensemble at time $i-1$. The forecast step is:
\begin{equation}
    \*x_i^{(k),\|f} = \!M_{i-1,  i}(\*x_{i-1}^{(k),\|a}) + \*\xi_i^{(k)},\quad k=1,\dots,N,
\end{equation}
where $\*\xi_i^{(k)}$ is a random draw from $ \!N(\*0,\*Q_i)$. 

The ETKF analysis step reads:
\begin{subequations}
\begin{eqnarray}
\label{eq:ETKF-analysis-anomalies}
    \*A^\|a_i &=& \*A^\|f_i\, \*T_i, \\
\label{eq:ETKF-analysis-mean}
    \bar{\*x}^\|a &=& \bar{\*x}^\|f + \*A^\|a_i\, \*Z^{\|a,\transp}\,\R^{-1}\,\*d_i,
\end{eqnarray}
\end{subequations}
where
\begin{subequations}
\begin{align}
\label{eq:transform-matrix}
\*T_i &= {\left(\Id - \*Z_i^{\|f,\transp}\,\S_i^{-1}\,\*Z_i^\|f\right)}^{\frac{1}{2}}, \\
\label{eq:S-matrix}
\S_i &= \*Z_i^\|f\,\*Z_i^{\|f,\transp} + \R_i, \\
\*A^\|f_i &= \frac{1}{\sqrt{N-1}}\left(\*X^{\|f}_i - \overline{\*x}^\|f_i\,\*1^\transp\right), \\
\*Z^\|f_i &= \frac{1}{\sqrt{N-1}}\left(\!H(\*\xf) - \overline{\!H(\*\xf)}\, \*1^\transp\right), \\
\*d_i &= \*y^\|o_i - \overline{\!H(\*X^{\|f}_i)},\\
\overline{\*x}^\|f_i &= \frac{1}{N}\sum_{k=1}^N \*X^{\|f,(k)}_i,\\
\overline{\!H(\*X^{\|f}_i)} &= \frac{1}{N}\sum_{k=1}^N \!H(\*X^{\|f,(k)}_i).
\end{align}
\end{subequations}
Here the unique symmetric square root of the matrix is used, as there is evidence of that option being the most numerically stable~\cite{sakov2008implications}.

The empirical forecast covariance estimate 
\begin{equation}
\label{eq:empirical-forecast-covariance}
\Ct{\xf_i}{\xf_i} = \*A^\|f_i\,\*A^{\|f,\transp}_i
\end{equation}
is inexact due to a multitude of deficiencies. One method to improve the empirical covariance estimate is inflation~\citep{anderson2001ensemble}, which is applied to the ensemble anomalies,
\begin{equation}
    \*A^\|f_i \leftarrow \alpha \, \*A^\|f_i,
\end{equation}
before any other computation is performed (meaning that it is also applied to the observation anomalies, $\*Z^\|f_i$ as well). The inflation parameter $\alpha>1$ is known to be a requirement for the EnKF to converge for linear models~\cite{popov2020explicit}.

\subsection{Covariance localization}
%
Traditional state-space localization of the empirical covariance \eqref{eq:empirical-forecast-covariance} is done by tapering, i.e., by using a Schur product of the empirical covariance with a localization matrix $\*\rho_i$:
\begin{equation}\label{eq:trad-loc}
 \*B^\|f_i =  \*\rho_i \circ \Ct{\xf_i}{\xf_i},
\end{equation}
where $\*\rho_i$ contains entries that are progressively smaller as the (physically-relevant) distance between the corresponding variables increases. 

The localized ETKF (LETKF)~\cite{hunt2007efficient} is an efficient implementation of localization for ETKF. The LETKF and its variants are considered to be one of the state-of-the-art EnKF methods. In the state-space approach to the LETKF, the $j$-th state space variable $\*x_{i,[j]}$ is assimilated independently of all others, with the observation space error covariance inverse replaced by
\begin{equation}\label{eq:rloc}
    \*R_i^{-1} \xleftarrow{} \*\rho_{i,[j]} \circ \*R_i^{-1},
\end{equation}
where $\*\rho_{i,[j]}$ is a diagonal matrix, with diagonal entries representing the decorrelation factors between all observation space variables and the $j$-th state space variable.  Each diagonal element represents a tapering factor, and is often chosen to be a function of the distance from the state-space variable$\*x_{i,[j]}$  being assimilated and the corresponding observation-space variable. The implicit assumption is that all observations are independent of each other, in both the observation error ($\*R_i$ is diagonal), and forecast error ($\*Z^\|f\*Z^{\|f,\Tp}$ is assumed to be diagonal).

\subsection{Covariance shrinkage}
In the statistical literature \cite{chen2009shrinkage,chen2010shrinkage,chen2011robust,ledoit2004well} covariance shrinkage refers to the methodology under which an empirical covariance is made to approach the ``true'' covariance from which the set of samples is derived. For the vast majority of statistical applications, there is no additional apriori knowledge about the distribution of the samples, thus assumptions such as Gaussianity and sphericity are made. In data assimilation applications, however, climatological estimates of covariance are commonplace.

Assume that one has access to a target covariance matrix $\*P$ that represents the \textit{a priori} knowledge about the error covariances. This matrix can be a climatological estimate of the covariance, or can be chosen through some ergodic assumption (with localization) using previous data. In existing data assimilation algorithms, such estimates most often exist for 4D-Var methods, and take the form of a static known background covariance that is an independent estimate from the current state.

We seek to combine this offline estimate of the covariance containing prior knowledge with the online estimate of the covariance obtained from the EnKF ensemble. In this paper we focus on an additive shrinkage covariance structure which is a linear combination of the  the target covariance matrix and the empirical covariance \eqref{eq:empirical-forecast-covariance}:
\begin{equation}
\label{eq:shrinkage-covariance}
    \*B^\|f_i = \gamma_i\,\mu_i\,\*P + (1-\gamma_i)\,\Ct{\xf_i}{\xf_i},
\end{equation}
with $\gamma_i$ represents the shrinkage factor (the linear combination coefficient) and $\mu_i$ represents a scaling factor. The choice of $\gamma_i$ is extremely important. In contemporary data assimilation literature (e.g., see \cite{asch2016data}) this factor is taken as a hyper-parameter whose optimal tuning could lead to significant reduction in error; however, the empirical tuning by trial and error is costly, and poor choices can offset possible improvements.

By employing a general invertible target matrix $\*P$, and optimizing for a 2-norm distance over the ``true'' covariance, a closed-form expression to compute the shrinkage factor $\gamma_i$ is proposed in  \cite{Stoica2008,Zhu2011}. In this derivation, weights are computed as follows:
\begin{equation}
    \gamma_i=\min\left(\frac{\frac{1}{N^2} \,\sum_{k=1}^N\norm{\*x_i^{(k),\|f}-\overline{\*x}^\|f_i}^4 -\frac{1}{N}\, \norm{ \Ct{\xf_i}{\xf_i}}^2}{\norm{\Ct{\xf_i}{\xf_i}-\*P}^2},1\right). \label{eq:KA}
\end{equation}
Since the estimate \eqref{eq:KA} is expensive to compute in an operational setting, here we will settle for a more computationally inexpensive method.
No assumptions about the structure of $\*P$ are made to compute $\gamma_i$. The general form \eqref{eq:shrinkage-covariance} can be reduced to a standard form where the target matrix is the (scaled) identity by defining: 
\begin{equation}
\label{eq:mu}
\*C_i \coloneqq \*P^{-\frac{1}{2}}\Ct{\xf_i}{\xf_i}\*P^{-\frac{1}{2}}, \qquad
    \mu_i = \frac{\tr(\*C_i)}{n},
\end{equation}
where the new target matrix $\mu_i\,\*I_{n \times n}$ represents a spherical climatological assumption on $\*C_i$. Equivalently we can write
\begin{equation}
\label{eq:shrinkage-covariance2}
    \*P^{-\frac{1}{2}}\*B^\|f_i\*P^{-\frac{1}{2}} = \gamma_i\,\mu_i\,\*I_{n \times n} + (1-\gamma_i)\,\*C_i.
\end{equation}
As is traditional with Kalman type methods, we make the assumption that all our samples are drawn from an underlying Gaussian distribution. This assumption allows for a simpler computation of $\gamma_i$.
The Rao-Blackwellized Ledoit-Wolf (RBLW) estimator~\cite{chen2009shrinkage} ~\cite[equation~(9)]{Sandu_2017_Covariance-Cholesky}:
\begin{equation}
\label{eq:RBLW}
\gamma_{i,\text{RBLW}} = \min \left[\vcenter{\hbox{$\displaystyle\frac{N - 2}{N(N+2)} + \frac{(n + 1)N - 2}{\hat{U}_i\,N(N+2)(n-1)}$}},\,\,\, 1\right],
\end{equation}
is the optimal estimate of the covariance shrinkage factor under Gaussian assumptions.  The computationally dominant (and interesting) term in \eqref{eq:RBLW} is the sphericity factor
\begin{equation}
\label{eqn:sphericity}
    \hat{U}_i = \frac{1}{n - 1}\left(\frac{n\, \tr(\*C_i^2)}{[\tr(\*C_i)]^2} - 1\right),
\end{equation}
which measures how similar the correlation structures of the sample and the target covariance are. For example if the both the target matrix $\*P$ and the empirical covariance matrix $\Ct{\xf_i}{\xf_i}$ are diagonal ($\*C_i$ is diagonal in \eqref{eq:shrinkage-covariance2}), then $\hat{U}_i=0$ in $\eqref{eqn:sphericity}$, meaning that the RBLW estimate \eqref{eq:RBLW} would be $\gamma_{i,\text{RBLW}}=1$. If, on the other hand, there is a large difference between the structures of $\*P$ and  $\Ct{\xf_i}{\xf_i}$ ($\*C_i$ has large off-diagonal elements in \eqref{eq:shrinkage-covariance2}), then the sphericity factor is close to $1$, forcing the RBLW estimate to be small (e.g., for $n=10^{10}$ and $N=50$, with $\hat{U}_i=1$, one has $\gamma_{i,\text{RBLW}} = 0.038$).

Note that if our samples are also used to calculate the sample mean, the effective sample size of the sample covariance is smaller by one, therefore for most practical applications one replaces $N$ by $N-1$ in \eqref{eq:RBLW}.

A drawback of the RBLW estimate is its reliance on the Gaussian assumption. A second drawback is that it is only valid for an over sampled ensemble with $N > n$, meaning that in the typical undersampled regime of EnKF with $N \ll n$, the factor is technically not well-defined. It is nonetheless still useful, in a similar fashion to ill-defined covariance estimates in the EnKF.

Aside from the inherent issues with the RBLW estimator, there are two major issues with its  application in the EnKF, both related to its reliance on the sphericity of $\*C_i$. First, when operating in the undersampled regime $N \ll n$, the estimate $\*C_i$ \eqref{eq:mu} is also undersampled, and the problem of ``spurious correlations'' will affect the measure of sphericity \eqref{eqn:sphericity}. The second related issue regards the climatological estimate $\*P$. Unless the climatological estimate accurately measures the correlation structure of the sample covariance, the shrinkage estimate \eqref{eq:shrinkage-covariance} could potentially not be representative of our current uncertainty. The long-term accuracy of the climatological estimate to the covariance is thus of great importance. 

Note that there are alternatives for non-invertible $\*P$. Commonly, a reduced spectral version of $\*P$ is known, $\*P = \!U\,\!L\,\!U^*$, with the $\!L$ being a diagonal matrix of $r\ll n$ spectral coefficients, and $\!U$ being an $n\times r$ matrix of orthonormal coefficients. The canonical symmetric pseudo-inverse square-root of $\*P$ would therefore be $\*P^{-\frac{1}{2}} = \!U\,\!L^{-\frac{1}{2}}\,\!U^*$. If $\sigma_k$ is the $k$-th singular value of $\*P^{-1/2}\*A^\|f$, then the traces appearing in \eqref{eqn:sphericity} can be computed as follows: 
\begin{gather*}
    \tr\left(\*C_i\right) = \sum_{k=1}^{N-1}\sigma_k^2,\qquad
    \tr\left(\*C_i^2\right) = \sum_{k=1}^{N-1}\sigma_k^4.
\end{gather*}
Note that only the first $N-1$ singular values are required for computation, even if $n \gg r \gg N-1$.

The choice of a suitable target matrix $\*P$ is very much an open question, and depends entirely on the problem at hand and on the available data. Some of the possible options include:
\begin{itemize}
    \item matrices that are used in variational data assimilation methods, 
    \item localized (through \eqref{eq:trad-loc}) estimates from historical data, such as from a previous cycle in quasi-periodic models, and
    \item estimates derived from more long-term models (such as climate models).
\end{itemize}
The above is a non-exhaustive list, and it would be up to the practitioner to decide the validity of one estimate over another. The mismatch of the target covariance with the covariance estimate derived from the dynamical ensemble through the sphericity factor~\eqref{eqn:sphericity} could also be used in an online manner to determine the utility of the target $\*P$, as a poor choice of the target matrix could significantly decrease the overall accuracy of the method, in the author's experience.

\section{ETKF implementation with stochastic shrinkage covariance estimates}

In ensemble-based methods our uncertainty is represented by an ensemble of samples of the underlying probability distribution. We wish to augment this representation of our uncertainty by augmenting the ensemble  of samples with historical (climatological) samples of said information, as the application of Bayes' rule requires that all available information  is used~\cite{jaynes2003probability}.

A naive approach to augmenting the ensemble would simply involve sampling from some known climatological distribution, for example sampling synthetic anomalies from a mean-zero Gaussian with known covariance, and appending this ensemble to our existing dynamical ensemble. This would, however, not be statistically sound, as the coupling between the two distributions would not be explicitly utilized. We therefore attempt to make use of the covariance shrinkage estimate \eqref{eq:shrinkage-covariance} to the covariance in order to couple the dynamical and synthetic ensembles correctly.

We build on previous work by Nino-Ruiz and Sandu ~\cite{Sandu_2015_covarianceShrinkage, Sandu_2019_Covariance-parallel} who proposed to replace the empirical covariance in EnKF with a shrinkage covariance estimator~\eqref{eq:shrinkage-covariance}. They showed that this considerably improves the analysis at a modest additional computational cost. Additional, it was shown that synthetic ensemble members drawn from a normal distribution with covariance $\*B^\|f$ are used to decrease the sampling errors.

In this work we develop an implementation of ETKF with a  stochastic shrinkage covariance estimator~\eqref{eq:shrinkage-covariance}. Rather than computing the covariance estimate ~\eqref{eq:shrinkage-covariance}, we build a synthetic ensemble by sampling directly from a distribution with covariance $\mu_i \*P$.
The anomalies of this synthetic ensemble are independent of the anomalies of the forecast EnKF ensemble.

Our approach works in a similar manner, but instead of simply augmenting the ensemble in a naive manner, we attempt to augment the ensemble in a statistically consistent manner by utilizing the theory behind optimal shrinkage estimators.
If the dynamical system is locally (in time) stationary, climatologies about the local time roughly describe a measure of averaged-in-space uncertainty. 


%
%

To be specific, let $\cxf \in \Re^{n \times M}$ be a synthetic ensemble with $M$ members (as opposed to the dynamic ensemble $\bxf_i$ with $N$ members) drawn from a climatological probability density. We denote the variables related to the synthetic ensemble by calligraphic letters. 

An important issue is the choice of the climatological distribution. As sampling from the dynamical manifold is impractical,  heuristic assumptions are made about the distributions involved. A useful known heuristic is the principle of maximum entropy (PME)~\cite{jaynes2003probability}. Assume that the mean and covariance of the distribution are known (through sampling), and that the distribution is supported over all of $\Re^n$. The synthetic ensemble  distribution of maximum entropy consistent with these assumptions is Gaussian:
\begin{equation}
\cxf_i \sim \mathcal{N}(\bar{\*x}^\|f_i,\mu_i\,\*P).
\end{equation}

The synthetic ensemble anomalies in the state and observation spaces are:
\begin{equation}
\label{eq:synthetic-anomalies}
\begin{split}
   \cxf_i &= \frac{1}{\sqrt{ M-1}}\left(\cxf_i - \overline{\!X}^\|f_i\;\*1_M^\transp\right) \in \Re^{n\times M},\\
    \!Z^\|f_i &= \frac{1}{\sqrt{ M-1}}\left(\!H(\cxf_i) - \overline{\!H(\cxf_i)}\;\*1_M^\transp\right) \in \Re^{m\times M}.
\end{split}
\end{equation}

The shrinkage  estimator \eqref{eq:shrinkage-covariance} of the forecast error covariance for $\*B^\|f_i$ is represented in terms of synthetic and forecast anomalies as follows:
\begin{equation}
\label{eq:Bf-shrinkage-stochastic}
\Bt^\|f_i = \gamma_i\,\!A^\|f_i\,\!A^{\|f,\transp}_i + (1-\gamma_i)\,\*A^\|f_i\,\*A^{\|f,\transp}_i.
\end{equation}
The Kalman filter formulation ~\cite{kalman1960new} yields the following analysis covariance matrix:
\begin{equation}
\label{eq:Ba-shrinkage-stochastic}
\Bt^\|a_i = \Bt^\|f_i - \Bt^\|f_i\,\*H_i^\transp\,\S_i^{-1}\,\*H_i \,\Bt^\|f_i,
\end{equation}
where $\S_i$ will be discussed later. 

Using the forecast error covariance estimate  \eqref{eq:Bf-shrinkage-stochastic} in \eqref{eq:Ba-shrinkage-stochastic} leads to the following analysis covariance:
%
%
\begin{equation}
\begin{split}
\Bt^\|a_i &= \gamma_i\,\!A^\|f_i\,\!A^{\|f,\transp}_i + (1-\gamma_i)\,\*A^\|f_i\,\*A^{\|f,\transp}_i \\
&\quad - \left(\gamma_i\,\!A^\|f_i\,\!Z^{\|f,\transp}_i + (1-\gamma_i)\,\*A^\|f_i\,\*Z^{\|f,\transp}_i\right)\,\S^{-1}_i\,\left(\gamma_i\,\!Z^\|f_i\,\!A^{\|f,\transp}_i + (1-\gamma_i)\,\*Z^\|f_i\,\*A^{\|f,\transp}_i\right),
\end{split}
\label{eq:goal-cov}\tag{goal-cov}
\end{equation}
%
%
which we refer to as the~ ``goal'' analysis covariance formula. Where in this paper the factor the factor $\gamma_{i}$ is chosen to be the RBLW estimate in \eqref{eq:RBLW}, unless otherwise specified.

The ensemble goal of our modified ensemble Kalman filter is to construct an $N$-member analysis ensemble such that the anomalies $\*A^\|a_i$ \eqref{eq:ETKF-analysis-anomalies} represent the~\eqref{eq:goal-cov} analysis covariance as well as possible:
\begin{equation}
\label{eq:goal-an}\tag{goal-an}
\textnormal{Find}~~\*A^\|a_i \in \Re^{n \times N}~~\textnormal{such that}:\quad
 \*A^\|a_i \,\*A^{\|a,\transp}_i \approx \Bt^\|a_i.
\end{equation}
In the proposed method, we enrich our forecast ensemble in a way that closely approximates the shrinkage covariance~\eqref{eq:shrinkage-covariance}.

\subsection{The stochastic shrinkage implementation}
We enrich the ensembles of forecast anomalies with synthetic anomalies \eqref{eq:synthetic-anomalies}:
\begin{equation}
\label{eq:type1-enriched-ensembles}
\begin{split}
    \#A^\|f_i &= \begin{bmatrix}\sqrt{1-\gamma_i}\,\*A^\|f_i &~~ \sqrt{\gamma_i}\,\!A^\|f_i\end{bmatrix} \in \Re^{n\times (N + M)},\\
    \#Z^\|f_i &= \begin{bmatrix}\sqrt{1-\gamma_i}\,\*Z^\|f_i &~~ \sqrt{\gamma_i}\,\!Z^\|f_i\end{bmatrix} \in \Re^{m\times (N + M)}.
\end{split}
\end{equation}
Next, we define a transform matrix \eqref{eq:ETKF-analysis-anomalies} that is applied to the enriched ensemble \eqref{eq:type1-enriched-ensembles}, and leads to an analysis ensemble that represents the target analysis covariance \eqref{eq:goal-an}. Specifically, we search for a transform matrix $\#T_i$ such that:
\begin{equation}
\label{eq:T1eq}
    \Bt^\|a_i = \#A^\|f_i\,\#T_i\,\#T_i^\transp\,\#A^{\|f,\transp}_i.
\end{equation}
Using the extended ensembles \eqref{eq:T1eq} the~\eqref{eq:goal-cov} becomes
\begin{equation}
\label{eq:t1ba}
    \Bt^\|a_i = \#A^\|f_i\,\left(\Id_{(N+M) \times (N+M)} - \#Z^{\|f,\transp}_i\,\S^{-1}_i\,\#Z^\|f_i\right)\,\#A^{\|f,\transp}_i,
\end{equation}
where, from \eqref{eq:S-matrix},
\begin{equation}\label{eq:shr-S-matrix}
    \S_i = \#Z^\|f_i\,\#Z^{\|f,\transp}_i + \R_i.
\end{equation}
The transform matrix \eqref{eq:transform-matrix} is a square root of ~\eqref{eq:T1eq}:
\begin{equation}
\label{eq:T-symmetric-sqrt}
    \#T_i = {\left(\Id_{(N+M) \times (N+M)} - \#Z^{\|f,\transp}_i\,\S^{-1}_i\,\#Z^\|f_i\right)}^{\frac{1}{2}}.
\end{equation}
We compute the analysis mean using the shrinkage covariance estimate. From  \eqref{eq:ETKF-analysis-mean}:
\begin{align}
\label{eq:type1-mean}
    \bar{\*x}^\|a_i &= \bar{\*x}^\|f_i + \#A^\|f_i\,\#T_i\,\#T_i^\transp\,\#Z^{\|f,\transp}_i\,\R^{-1}_i\,\*d_i,
\end{align}
where the full analysis covariance  estimate ~\eqref{eq:t1ba} is used.
In addition, we achieve the \eqref{eq:goal-an} by keeping the first $N$ members of the transformed extended ensemble, or equivalently, the first $N$ columns of the symmetric square root \eqref{eq:T-symmetric-sqrt}.  From \eqref{eq:ETKF-analysis-anomalies} we have:
\begin{align}
\label{eq:type1-anomalies}
    \*A^\|a_i &= \frac{1}{\sqrt{1-\gamma_i}}\,\bigl[\#A^\|f_i\,\#T_i\bigr]_{:,1:N} = \frac{1}{\sqrt{1-\gamma_i}} \,\#A^\|f_i\,\breve{\#T}_i,
    \quad \breve{\#T}_i = \left[\#T_i\right]_{:,1:N}.
\end{align}
An alternative approach to achieve the \eqref{eq:goal-an} is to look for a low-rank, approximate square root instead of the symmetric square root \eqref{eq:T-symmetric-sqrt}. Specifically, we seek a transformation matrix $\widehat{\#T}_i$ such that:
\begin{equation}
\label{eq:T-lowrank-sqrt}
\widehat{\#T}_i \in  \Re^{(N+M) \times N}, \qquad
    \widehat{\#T}_i\,\widehat{\#T}_i^T \approx \Id_{(N+M) \times (N+M)} - \#Z^{\|f,\transp}_i\S^{-1}_i\#Z^\|f_i.
\end{equation}
The calculation of the symmetric square root \eqref{eq:T-symmetric-sqrt} requires an SVD of the right hand side matrix. With the same computational effort one can compute the low rank transformation:
\begin{equation}
\label{eq:T-symmetric-sqrt2}
\begin{split}
\mathbf{U}\, \boldsymbol{\Sigma} \, \mathbf{U}^T &= {\left(\Id - \#Z^{\|f,\transp}_i\S^{-1}_i\#Z^\|f_i\right)}, \qquad 
\mathbf{U}, \boldsymbol{\Sigma} \in  \Re^{(N+M) \times (N+M)};\\
    \breve{\#T}_i &= \mathbf{U}\, \boldsymbol{\Sigma}^{1/2} \, [\mathbf{U}_{1:N,:}]^T \in  \Re^{(N+M) \times N} \qquad\textnormal{(symmetric square root \eqref{eq:T-symmetric-sqrt})}; \\
    \widehat{\#T}_i &= \mathbf{U}_{:,1:N}\, \boldsymbol{\Sigma}^{1/2}_{1:N,1:N} \in  \Re^{(N+M) \times N} \qquad \textnormal{(low rank square root \eqref{eq:T-lowrank-sqrt})}.
\end{split}
\end{equation}
The mean calculation \eqref{eq:type1-mean} is the same. The ensemble transform produces $N$ transformed ensemble members that contain ``mixed'' information from both the physical and the synthetic ensembles:
\begin{align*}
\label{eq:type2-anomalies}
    \*A^\|a_i &= \frac{1}{\sqrt{1-\gamma_i}} \,\#A^\|f_i\,\widehat{\#T}_i.
\end{align*}

\subsection{Localization}
\label{sec:localized-stochastic-shrinkage}
It is possible to combine the proposed stochastic shrinkage approach with traditional localization. The LETKF implementation~\cite{hunt2007efficient} computes transform matrices for subsets of variables, corresponding to localized spatial domains. In a similar vein one can combine our shrinkage algorithm with classical localization, as follows. Subsets of variables of the enriched ensembles \eqref{eq:type1-enriched-ensembles} are used to compute local transform matrices \eqref{eq:T-symmetric-sqrt} or \eqref{eq:T-lowrank-sqrt}, which are then applied to transform the corresponding local subsets, i.e. to compute the corresponding rows in equations \eqref{eq:type1-anomalies} or \eqref{eq:type1-anomalies}, respectively.
Utilizing the Sherman-Morrison-Woodbury identity~\cite{petersen2008matrix}, it is possible to decompose the inverse of \eqref{eq:shr-S-matrix} into,
\begin{equation}
    \*S_i^{-1} = \*R_i^{-1} - \*R_i^{-1}\!Z_i\left(\!Z_i^T\*R_i^{-1}\!Z_i + \*I_{N + M}\right)\!Z_i^T\*R_i^{-1},
\end{equation}
where the inverse observation covariance is replaced by the localized variant in \eqref{eq:rloc}.

\section{Numerical experiments}

In the numerical experiments we aim to assess the performance of the methodology in three different regimes: (i) a small scale model (Lorenz '96) to empirically test the performance of the optimally estimate the covariance shrinkage factors $\gamma$ \eqref{eq:RBLW} constants against hand-picked values, (ii) a medium scale model (Quasi-geostrophic equations) with small observation errors to test the unlocalized shrinkage covariance methodology against the state-of-the-art LETKF, (iii) a geophysical model (shallow water on the sphere)  with large observation errors to test our localized methodology against the LETKF.

All test problem implementations are available in the `ODE Test Problems' suite \cite{otp, otpsoft}.

\subsection{The Lorenz'96 model  (L96)}


We first consider the 40-variable Lorenz '96 problem~\cite{lorenz1996predictability},
\begin{equation}
\label{eq:Lorenz}
    \left[y\right]'_i = - \left[y\right]_{i-1}\left(\left[y\right]_{i-2} - \left[y\right]_{i+1}\right) - \left[y\right]_i + F, \quad i=1,\dots,40, \quad F=8.
\end{equation}
Assuming \eqref{eq:Lorenz} is ergodic (thus having a constant spatio-temporal measure of uncertainty on the manifold of the attractor), we compute the target covariance matrix $\*P$ as the empirical covariance from $10,000$ independent ensemble members run over $225$ days in the system (where $0.05$ time units corresponds to 6 hours), with an interval of 6 hours between snapshots. This system is known to have 13 positive Lyapunov exponents, with a Kaplan-Yorke dimension of about 27.1 \cite{popov2019bayesian}. 

The time between consecutive assimilation steps is $\Delta t = 0.05$ units, corresponding to six hours in the system. All variables are observed directly with an observation error covariance matrix of $\R_i = \Id_{40}$. The time integration of the model is performed with RK4 the fourth order Runge-Kutta scheme RK4 \cite{hairer1991solving} with a step size $h = \Delta t$. The problem is run over 2200 assimilation steps. The first 200 are discarded to account for model spinup. Twenty independent model realizations are performed in order to glean statistical information thereof.

\subsection{L96 assimilation results}

\begin{figure*}[t]
    \centering
    \includegraphics[width=\linewidth]{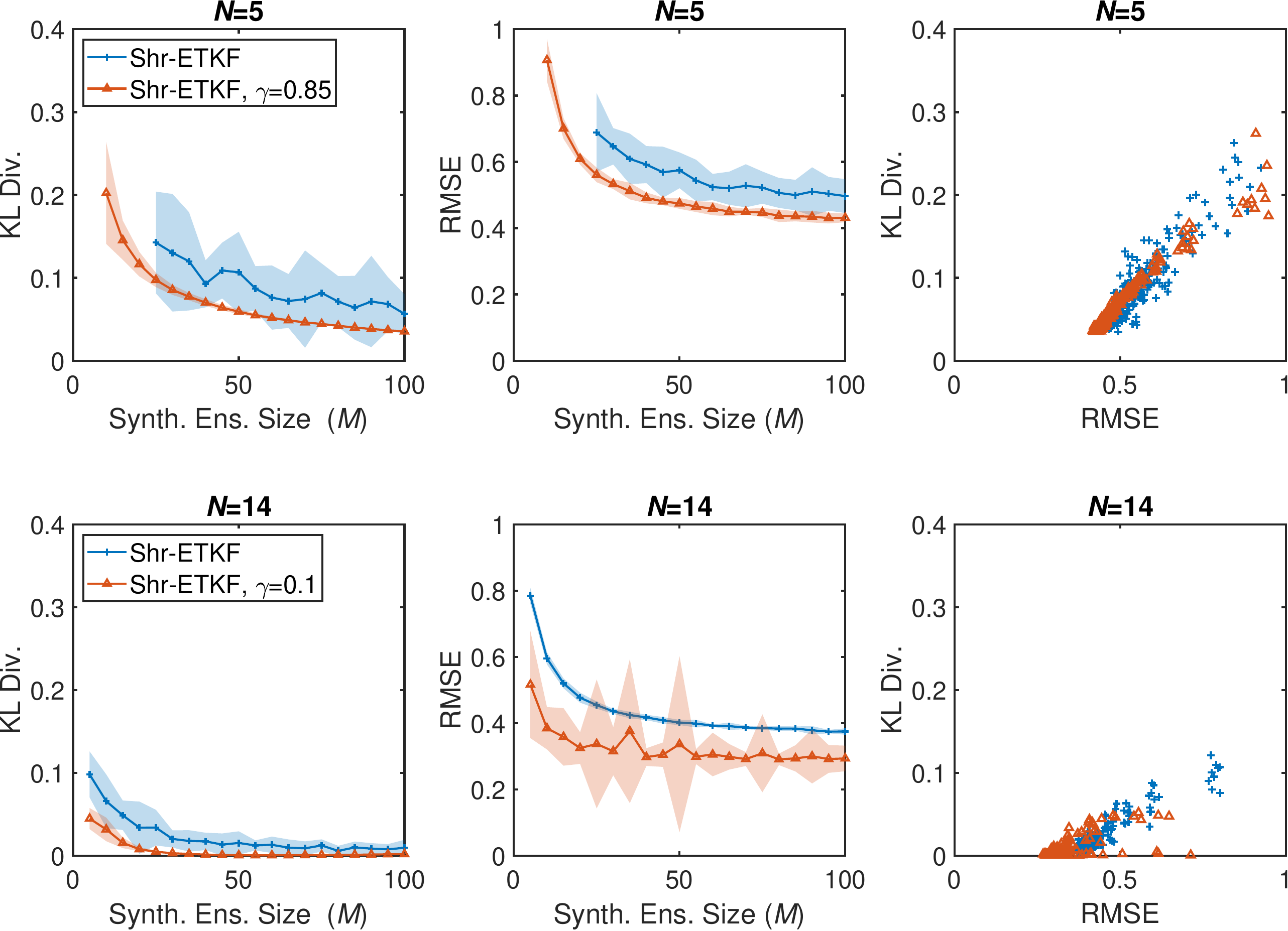}
    \caption{Results for the L96 problem with dynamic ensembles sizes of $N=5$ and $N=14$, inflation factor $\alpha=1.1$, and different synthetic ensemble sizes $M$. We compute the KL divergence of the rank histogram~\eqref{eq:rank-histogram} and the RMSE~\eqref{eq:rmse} for the methods. Error bars show two standard deviations.}\label{fig:KLRMSEM}
\end{figure*}

\begin{figure*}[t]
    \centering
    \includegraphics[width=0.66\linewidth]{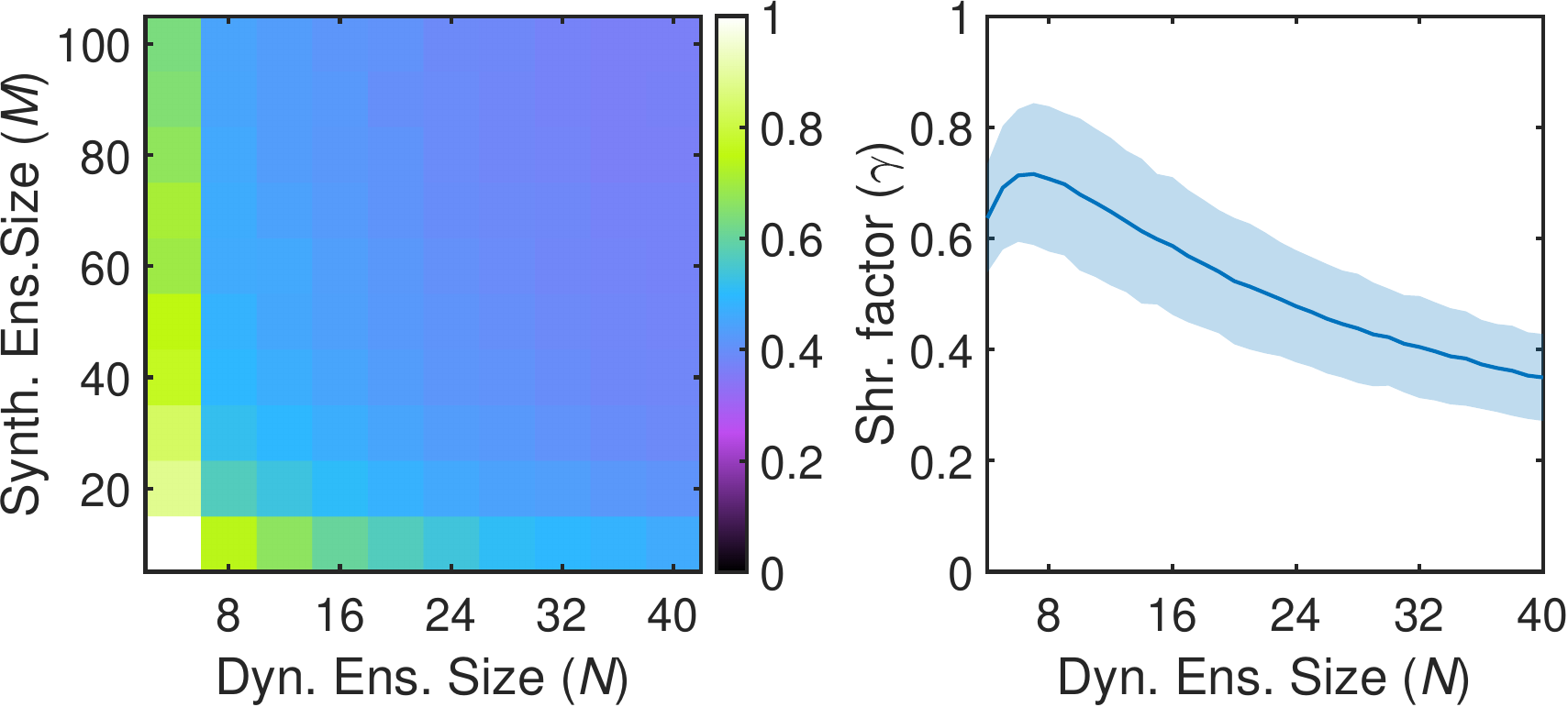}
    \caption{Results for the L96 problem. The left panel presents the analysis RMSE for various values of the dynamic and synthetic ensemble sizes. The right panel presents the shrinkage factor $\gamma$ \eqref{eq:RBLW} for a synthetic ensemble size of $M=100$ over a number of assimilation steps, with error bars showing two standard deviations.}
    \label{fig:l96-M-versus-N-versus-G}
\end{figure*}

We assess the quality of the analysis ensembles using a rank histogram~\cite{hamill2001interpretation}, cumulative over 20 independent runs. For a quantitative metric we compute the KL divergence from $Q$ to $P$,
\begin{equation}\label{eq:rank-histogram}
    D_{KL}\left(P\middle|\middle|Q\right) = -\sum_k P_i\log\left(\frac{P_k}{Q_k}\right),
\end{equation}
where $P$ is the uniform distribution and $Q$ is our ensemble rank histogram, and $P_k$ \& $Q_k$ are the discrete probabilites associated with each bin. A low KL divergence would indicate that our rank histogram is close to uniform, and thus the ensemble is representative of the truth.

Additionally, for testing the accuracy of all our methods we compute the spatio-temporal analysis RMSE,
\begin{equation}\label{eq:rmse}
    \sqrt{\frac{1}{K n}\sum_{i=1}^K \sum_{j=1}^n \left[x_i^\|a - x_i^\|t\right]_j^2},
\end{equation}
with $K$ representing the amount of snapshots at which the analysis is computed, and $\left[x_i\right]_j$ is the $j$th component of the state variable at time $i$.

For the given settings of a severely undersampled ensemble ($N=5$) and mild inflation ($\alpha = 1.1$), we compare the Gaussian sampling methodology coupled to the RBLW formulation for the shrinkage factor $\gamma$~\eqref{eq:RBLW}, with the optimal static $\gamma=0.85$ shrinkage factor. For a dynamic ensemble that captures the positive error growth modes ($N=14$) will will compare the RBLW estimator with the optimal static $\gamma=0.1$. We will compare the mean and variance of the KL divergence of the rank histogram of the variable $[y]_{17}$ from the uniform, and the statistics of the spatio-temporal RMSE.

The results are reported in Figure~\ref{fig:KLRMSEM}. For both an undersampled and sufficient ensemble, the optimal shrinkage factor has a smaller mean error, and a smaller KL divergence with less variance (top left, top middle, bottom left, and bottom middle panels). In the undersampled case, the RBLW estimator induces more variance into the RMSE (top middle panel). For the sufficiently sampled ensemble, however, the optimal static shrinkage value induced significantly more variance into the error, with the RBLW estimator reducing the error significantly (bottom middle panel). 

It is possible that a better estimator than RBLW may get the `best of both worlds' and induce low error with low variance, though this is as-of-yet out of reach. This is to be expected as the RBLW estimate is only accurate in the limit of ensemble size, and there is no theory about its accuracy in the undersampled case. In the authors' experience other estimators such as OAS, while having the theoretically desired properties, perform empirically worse in conjunction with ensemble methods. Currently, for a modest reduction in accuracy, one of the hyperparameters can be {estimated online} by the methodology.

For the second round of experiments with Lorenz '96, reported in Figure~\ref{fig:l96-M-versus-N-versus-G}, we compare analysis errors when the synthetic and dynamic ensemble sizes are modified (left panel). It is evident that increases in both dynamic and synthetic ensemble size lead to lower error. In the right panel we also compare dynamic ensemble size to  the values of $\gamma$ that are produced. It is clear that an increase in dynamic ensemble size decreases the need for shrinkage.

\subsection{The Quasi-Geostrophic model (QG)}

\begin{figure*}[t]
    \centering
    \includegraphics[width=0.6666\linewidth]{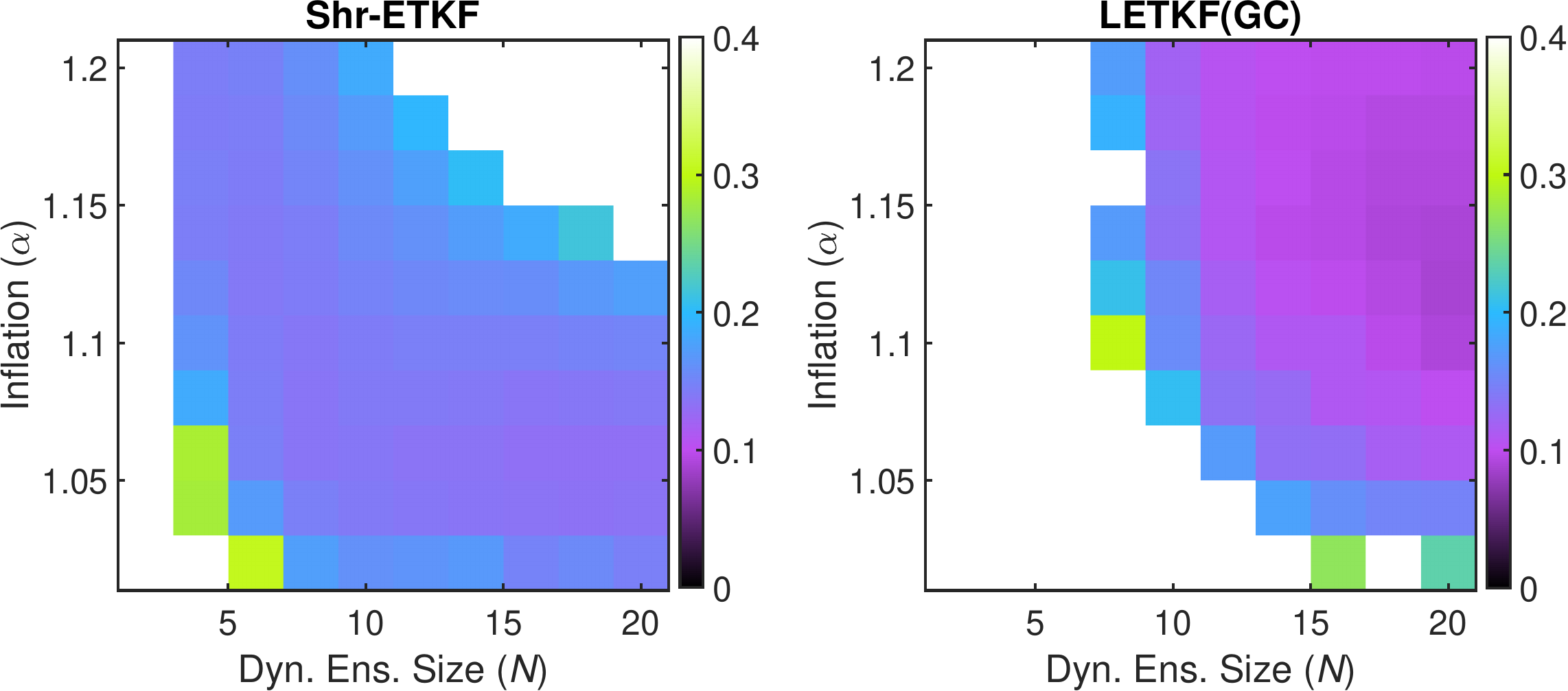}
    
    \begin{minipage}[t]{.33333\linewidth}
    \centering
    \end{minipage}%
    \begin{minipage}[t]{.33333\linewidth}
    \centering
    \end{minipage}%
    \begin{minipage}[t]{.33333\linewidth}
    \centering
    \end{minipage}
    
    \caption{Analysis RMSE results for the QG model. The experiments use a synthetic ensemble size $M=100$ and Gaussian samples. Results are compared against LETKF with the Gaspari-Cohn decorrelation function (GC)}
    \label{fig:qgT1T2}
\end{figure*}

We follow the QG formulations given in~\cite{san2015stabilized,mou2019data}. We discretize the equation
\begin{equation}
  \begin{split}
    \label{eq:QG}
    \omega_t + J(\psi,\omega) - {Ro}^{-1}\, \psi_x &= {Re}^{-1}\, \Delta\omega + {Ro}^{-1}\,F, \\
    J(\psi,\omega)&\equiv \psi_y\, \omega_x - \psi_x\, \omega_y,
  \end{split}
\end{equation}
where $\omega$ stands for the vorticity, $\psi$ stands for the stream function, $Re$ is the Reynolds number, $Ro$ is the Rossby number, $J$ is the Jacobian term, and $F$ is a constant (in time) forcing term.

The relationship between stream and vorticity, $\omega = -\Delta\psi$ is explicitly enforced in the evaluation of the ODE. The forcing term is a symmetric double gyre,
\begin{equation}
  F = \sin\left(\pi(y-1)\right).
\end{equation}
Homogeneous Dirichlet boundary conditions are enforced on the spatial domain $[0,1]\times[0,2]$. The spatial discretization is a second order central finite difference for the first derivatives, and the Laplacian, with the Arakawa  approximation \cite{arakawa1966computational} (a pseudo finite element scheme~\cite{jespersen1974arakawa})  used for computing the Jacobian term. All spatial discretizations exclude the trivial boundary points from explicit computation.

The matrix $\*P$ is approximated from 700 snapshots of the solution about 283 hours apart each, with Gaspari-Cohn localization applied, so as to keep the matrix sparse. The true model is run outside of time of the snapshots so as to not pollute the results. Nature utilizes a $255\times511$ spatial discretization, and the model a $63\times127$ spatial discretization. Observations are first relaxed into the model space via multigridding~\cite{zubair2009efficient}, 
then 150 distinct spatial points (using an observation operator similar to \cite{sakov2008deterministic}) from the non-linear observation operator,
\begin{equation}
    \!H(\psi) = \sqrt{\psi_x^2 + \psi_y^2},
\end{equation}
representing zonal wind magnitude, are taken. The observation error is unbiased, with covariance $\*R = 4\Id_{150}$. The number of synthetic ensemble members is fixed at a constant $ M = 100$, as to be more than the number of full model run ensemble members, but significantly less than the rank of the covariance.
Observations are taken $\Delta t=0.010886$ time units (representing one day in model space) apart.
We run a total of 350 assimilation steps, taking the first 50 as spinup. Results are averaged over 5 model runs (with the same nature run, but different initializations of the dynamic ensemble), with diverging runs treated as de-facto infinite error.

\subsection{QG assimilation results}
Figure~\ref{fig:qgT1T2} reports the results with the QG model. Comparing our methodology to the LETKF with an optimally tuned Gaspari-Cohn (GC)~\cite{gaspari1999construction} localization (such that both error and stability are prioritized), we see that GC significantly decreases the error for larger values of $N$ and $\alpha$, but is not stable for more operational under-sampled dynamic ensemble sizes and low inflation factors, as opposed to our shrinkage method. Possible sources of error are both the nonlinear observations and the coarse approximation to the covariance estimate. 

These results lend additional support to the argument that shrinkage alone is not enough. Localization is still required in operational settings, and combining both might yield a positive result.

The quasi-geostrophic results indicate that our methodology holds promise to be of use for practical data assimilation systems, and that the methodology can handle observations that are non-linear transformations of the state representation. However, the methodology needs to be refined with more optimal shrinkage factors for operational undersampled empirical covariances.

An operational implementation of the LETKF requires $m\times N$ linear solves and $m$ matrix square roots, while our stochastic shrinkage algorithm requires $N + M$ linear solves and one matrix square root. Thus as the number of observations grows, the stochastic shrinkage methodology becomes a lot more compelling.

\subsection{Shallow water on a sphere (SWS)}

\begin{figure*}[t]
    \centering
    \includegraphics[width=0.666\linewidth]{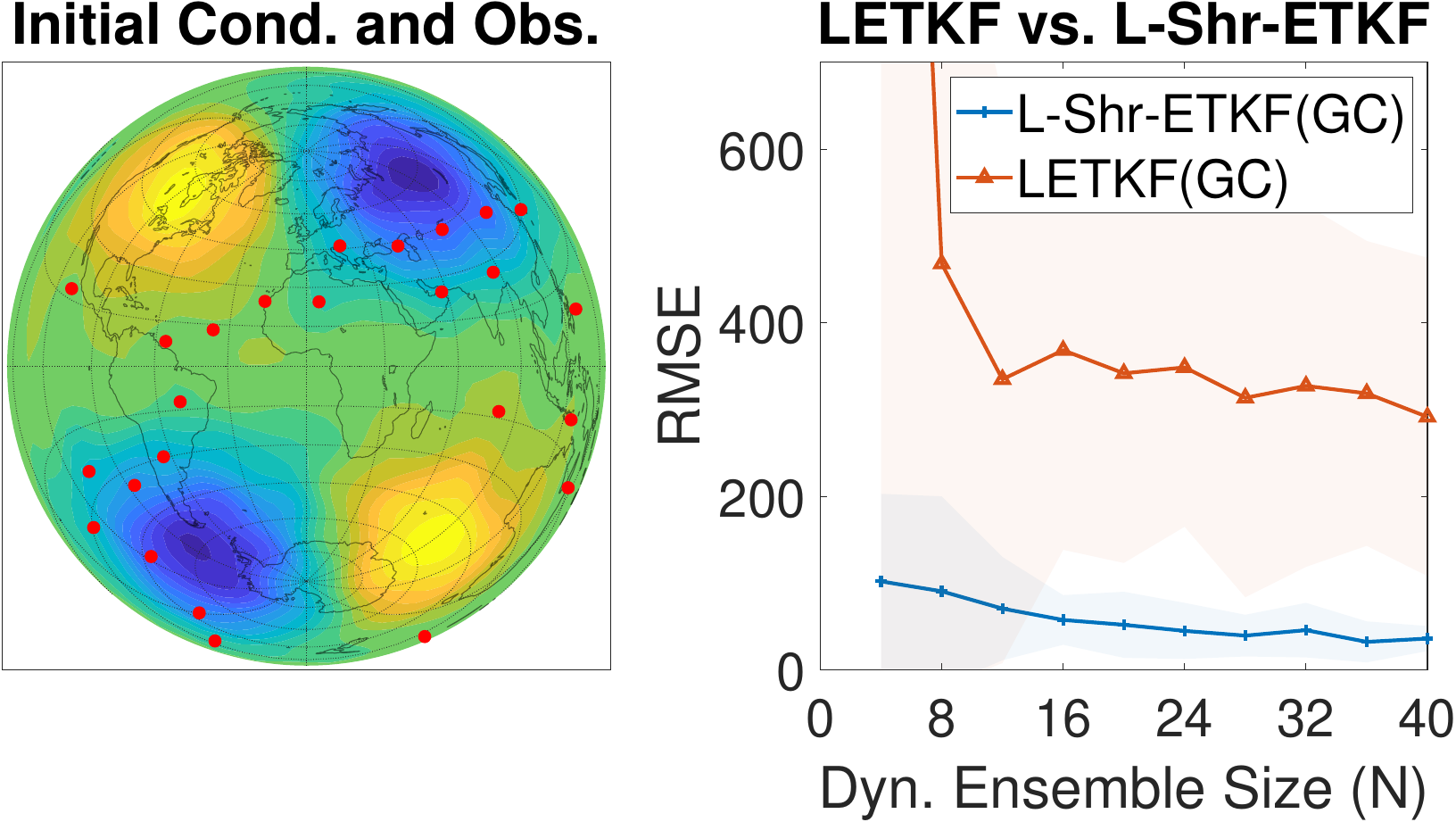}
    
    \caption{Left panel: initial condition of the water height with blue represented lower than average and yellow representing higher than average, and observation locations (red points). Right panel: analysis RMSE for the localized shrinkage ETKF, and the localized ETKF with the Gaspari-Cohn decorrelation function, with the error bars representing two standard deviations.}
    \label{fig:shallow-water-experiments}
\end{figure*}

The last round of experiments aims at validating the Localized Shrinkage ETKF on a different geophysical problem of interest.
To that end we employ the shallow water equations~\cite{flyer2009radial,neta1997analysis} on the sphere, which represent an approximation of the atmospheric dynamic over Earth.
We use a modification of the Cartesian shallow water equations,
\begin{align}
    h_t &= -\nabla\cdot(h\cdot \*u),\\
    \*u_t &= - (\*u\cdot \nabla)\*u - f(\*p\times\*u) - g\nabla h,
\end{align}
under the constraint that the flow is confined to a spherical approximation to the Earth; the radius of the sphere is one spatial unit. Here, $f$ is the Coriolis force, $g$ is gravity,  $h$ is the height of the water, and $\*p$ and $\*u$ are the $x$, $y$, and $z$ positions and velocities, respectively. We follow the radial basis function formulation in~\cite{flyer2009radial} for the spatial discretization with $100$ points, for a total state space dimension of $n=400$. We take the order three Buhmann function~\cite{buhmann1998radial} with a Cartesian radius of $r=2$ on the unit sphere (representing full coverage). We use a third order adaptive strong stability preserving method~\cite{macdonald2003constructing} for time integration of this system.

We observe the height at ten locations over the domain; the velocities are unobserved. Observations are taken every $\Delta t = 1$ day over the assimilation window. The observation covariance is $\*R = 100\,\*I_{10}$, to simulate a noisy observation scenario.

We compare the localized variant of the stochastic shrinkage approach (see section~\ref{sec:localized-stochastic-shrinkage}), which we term the L-Shr-ETKF, against LETKF. We select a synthetic ensemble size $M=250$. 
For localization, we use a great circle  radius of $r = \pi/5$ spatial units, with the Gaspari-Cohn decorrelation function, as this was found to be approximately optimal for the LETKF by manual tuning. The best inflation factors obtained by manual tuning are used, as follows: $\alpha = 1.001$, and $\alpha = 1.05$ for  LETKF. 

We run a total of three months of observations for January, February and March (90 days), discarding the first 31 days of January as spinup, and observing the analysis RMSE for a range of physical ensemble sizes, $N \in [4, 40]$. Twelve total independent runs are taken to account for possible spurious results.

\subsection{SWS Results}

The left panel of figure~\ref{fig:shallow-water-experiments} show the initial conditions, and observations for the shallow water equations. The initial conditions were chosen to be quasi-stable so that they would slowly diverge from a cyclic solution.
The right panel shows the results comparing  L-Shr-ETKF against the state-of-the-art LETKF. It can be clearly seen that even for large ensemble sizes up to 40, the LETKF error is higher than the observation error of $100$. The LETKF also suffers from large variability in the error from various different initial ensembles.
L-Shr-ETKF, on the other hand, matches the observation error for an ensemble size of $N=4$ dynamical members, and has lower error than that of the observations for all  larger dynamical ensemble sizes. The variability of the L-Shr-ETKF error is also substantially smaller than that of LETKF.

The results clearly demonstrate that, in a small-ensemble high-observation-error regime where LETKF performs relatively poorly, the proposed L-Shr-ETKF algorithm provides robust analyses.

\section{Discussion}

Shrinkage covariance matrix estimators were shown to greatly improve the performance of the EnKF ~\cite{Sandu_2015_covarianceShrinkage}.
This work extends the the idea of covariance shrinkage to the ensemble transform Kalman filter. Instead of enhancing the covariance estimate, we propose enhancing the ensemble with a synthetic ensemble derived from the target matrix of the shrinkage approach. By applying the ETKF formulas to this enhanced ensemble, we develop the Shr-ETKF, whose internal representation of the Kalman gain is approximately based on the shrinkage estimate of the covariance.

We compare Shr-ETKF to the current state-of-the-art LETKF algorithm on several test problems. Lorenz '96 model results indicate that the new filter performs worse in the under-sampled regime than the best `static' shrinkage method, and performs better (in terms of less variance in the error) than an optimal dynamic shrinkage method for the sufficiently sampled case. Results with QG model indicate that our method could potentially be used to augment operational LETKF implementations, but not in the low-observation-error regime.
Results with the shallow-water equation on a sphere model show that a localized stochastic covariance shrinkage ETKF can perform significantly better than the LETKF in a high-observation-error regime.  

These results indicate that L-Shr-ETKF can be potentially utilized in an operational framework to improve the performance of LETKF while keeping the dynamical ensemble size (the number of forecast model runs) small. Additional work is needed to devise better heuristic estimates of the shrinkage factor $\gamma$.

\begin{acknowledgments}
The first two authors would like to thank Traian Iliescu and Changhong Mou for their in-depth help with understanding of the Quasi-Geostrophic model, Steven Roberts, the primary maintainer of the ODE Test Problems package, Amit N Subrahmanya for his help with the shallow water on a sphere problem, and the rest of the members of the Computational Science Laboratory at Virginia Tech for their continuous support. 

The authors would like to thank the four previous referees for their  insightful feedback that lead to an improved paper.

The first two authors were supported, in part, by the National Science Foundation through awards NSF ACI--1709727, NSF CCF--1613905, NSF CDS\&E--MSS 1953113, AFOSR through the award AFOSR DDDAS 15RT1037, and by DOE ASCR through the award  DE--SC0021313.

The last author was supported by the Research Council of Norway and the companies AkerBP, Wintershall--DEA, V{\aa}r Energy, Petrobras, Equinor, Lundin and
Neptune Energy, through the Petromaks--2 project (280473) DIGIRES (\url{http://digires.no})
\end{acknowledgments}

\section*{References}
\bibliography{biblio}

\end{document}